\documentclass[useAMS,usenatbib]{mn2e}
\usepackage{times}
\usepackage{graphics,epsfig}
\usepackage{graphicx}
\usepackage{amssymb}

\title[Turbulence in supernova remnants]{Magnetohydrodynamic turbulence in 
supernova remnants}

\author[N. Roy et al.]{Nirupam Roy $^{1}$\thanks{E-mail: nirupam@ncra.tifr.res.in~(NR); somnath@cts.iitkgp.ernet.in (SB); prasun@cts.iitkgp.ernet.in~(PD); chengalu@ncra.tifr.res.in~(JNC)}, Somnath Bharadwaj $^{2}$\footnotemark[1], Prasun Dutta $^{2}$\footnotemark[1] and Jayaram N. Chengalur $^{1}$\footnotemark[1]\\
       $^{1}$NCRA-TIFR, Post Bag 3, Ganeshkhind, Pune 411 007, India\\
       $^{2}$Department of Physics and Meteorology \& Centre for Theoretical Studies, IIT Kharagpur, Kharagpur 721 302, India}

\begin{document}
\date{Accepted yyyy month dd. Received yyyy month dd; in original form yyyy 
month dd}

\pagerange{\pageref{firstpage}--\pageref{lastpage}} \pubyear{2008}

\maketitle

\label{firstpage}

\begin{abstract}

We present estimates of the angular power spectra of the synchrotron radiation 
intensity fluctuations at 6 and 20 cm for the shell type supernova remnant Cas 
A and the filled-centre Crab supernova remnant. We find that the intensity 
fluctuations of both sources have a power law power spectrum with index 
$-3.24\pm0.03$. This power law power spectrum is consistent with the 
magnetohydrodynamic turbulence in the synchrotron emitting plasma. For Cas A, 
there is a break in the power spectrum and the power law index changes from 
$-3.2$ to $-2.2$ at large angular scale. This transition occurs at an angular 
scale that corresponds to the shell thickness of Cas A. We interpret this as a 
transition from three dimensional turbulence to two dimensional turbulence on 
scales that are respectively smaller and larger than the shell thickness.
 
\end{abstract}

\begin{keywords}
MHD --- ISM: general --- ISM: individual (Cas A, Crab Nebula) --- supernova remnants --- turbulence 
\end{keywords}

\section{Introduction}
\label{sec:int}

Supernovae and supernova remnants play a very important role in astrophysics 
at the galactic scale. They cause the heating of the interstellar medium, 
acceleration of cosmic rays and enrichment of the interstellar medium with 
heavy elements created in the stellar core or in the supernova explosion. The 
shock wave traveling through the interstellar medium may also cause the gas 
clouds to collapse to form new stars. Thus, they work as a link between the 
gaseous and stellar components of the Galaxy. Based on their large scale 
structure, supernova remnants are broadly classified into three types: 
shell-type remnants, filled-centre remnants (or plerions) and composite 
remnants \citep{ws88}. In addition to the large scale shell-like or filled-centre 
structures, all these remnants show a very rich and complicated structure over 
a wide range of scale and frequency of observation.

Although there have been many high resolution and high sensitivity 
multiwavelength observations of Galactic supernova remnants, there has not 
been, to the best of our knowledge, any systematic study to quantify the fine 
scale structure. Here we present 6 and 20 cm observations and estimates of the 
angular power spectra of the intensity fluctuation over a wide range of 
angular scales for the supernova remnants Cas A and the Crab Nebula. 

The Crab Nebula (G184.6$-$5.8) is a supernova remnant and pulsar wind nebula 
in the constellation of Taurus in the third Galactic quadrant. It is a 
filled-centre nebula, $7\arcmin\times5\arcmin$ in size \citep{van70} at a 
distance of approximately $2$ kpc \citep{tr73}. The remnant is of the famous 
supernova of 1054 AD and the Crab pulsar is observed to be at the centre of 
this nebula. At radio wavelengths the source is quite strong with a flux 
density of $1040$ Jy at $1$ GHz and, beside its filled centre structure, shows 
faint jet or tube like extension from the north edge of the remnant. 
Cassiopeia A or Cas A (G111.7$-$2.1) is in the constellation Cassiopeia in the 
second Galactic quadrant. This is a shell type supernova remnant of diameter 
$5\arcmin$ at a distance of approximately $3.4$ kpc \citep{re95}. The shell 
thickness estimated from the radial brightness profile at radio wavelengths is 
found to be approximately $30\arcsec$. At these wavelengths, it shows a clear 
shell like structure with compact emission knots and is one of the strongest 
radio sources in the sky with a flux density of $2720$ Jy at $1$ GHz. It is 
most probably the remnant of a late 17th century supernova \citep{fe06}. There 
is some spectroscopic evidence that Cas A was a type IIb supernova 
\citep{kr08}. More details of both these supernova remnants can be found in 
the Galactic supernova remnants catalogue \citep{gr04}\footnote[1]{See 
http://www.mrao.cam.ac.uk/surveys/snrs/ for an updated version.}.

At radio wavelengths, the dominant contribution to the supernova remnants' 
emission comes from the synchrotron radiation emitted by the relativistic 
electrons in the presence of magnetic fields. The observed structures over a 
wide range of scales are most probably result of the magnetohydrodynamic 
turbulence in the emitting plasma. Since the synchrotron radiation intensity 
fluctuation will depend on fluctuation of both density and magnetic field 
strength, it is expected that the power spectrum will reveal interesting 
information about the density and magnetic field fluctuation as well as about 
the nature of the turbulence in the plasma.

We present here the estimates of the power spectrum obtained directly from the 
interferometric measurements of the visibility function of the sources. The 
analysis technique is briefly described below in \S\ref{sec:atech}. In 
\S\ref{sec:dar}, the details of the observational data and the results are 
given. Finally, we summarize and present our conclusions in \S\ref{sec:con}.

\section{Analysis technique}
\label{sec:atech}

Assuming that the angular extent of the source is small, the angular power 
spectrum, $P(u,v)$, of the intensity fluctuation of synchrotron radiation 
$\delta I(l,m)$ can be written as 
\begin{equation}
P(u,v)=\int\int\xi(l, m)e^{-2\pi i(ul+vm)}dldm
\end{equation}
where $(l,m)$ is the direction on the sky, $(u,v)$ is the inverse angular 
separations and $\xi$ is the autocorrelation function of the intensity 
fluctuation
\begin{equation}
\xi(l-l^{\prime}, m-m^{\prime})=\langle\delta I(l,m)\delta I(l^{\prime}, 
m^{\prime})\rangle.
\end{equation}
Here the angular brackets imply an average across different positions and 
directions on the sky. If the angular extent of the source is not small enough 
then, instead of taking the Fourier transform, a spherical harmonic 
decomposition of the autocorrelation function is to be done to get the angular 
power spectrum. Throughout this analysis, it is assumed that the angular size 
of the source is small and the statistical properties of the small scale 
intensity fluctuations are homogeneous and isotropic. Hence, the intensity 
fluctuation power spectrum $P(u,v)$ is a function of the magnitude $U = 
\sqrt{u^2+v^2}$ only and is independent of the direction.

Since the complex visibility function $V(u,v)$ measured by an interferometer 
is the Fourier transform of the source brightness distribution $I(l,m)$,
\begin{equation}
V_s(u,v)=\int\int I(l, m)e^{-2\pi i(ul+vm)}dldm
\end{equation}
where $(u,v)$ is the baseline or the projected antenna separation in units of 
the wavelength of observation and is associated with an inverse angular scale, 
one can estimate the angular power spectrum directly from the measured 
visibility function. It can be easily shown that the squared modulus of the 
visibility is a direct estimator of the intensity fluctuation power spectrum
\begin{equation}
P(u,v)=\langle V_s(u,v)V^*_s(u,v)\rangle
\end{equation}
where the angular brackets denote an average over all possible orientations of 
the baselines. 

This method for estimating the power spectrum from the complex visibility 
function has been used earlier by \citet{cr83} and \citet{gr93}. The technique 
of direct visibility based estimation of power spectrum has also been used and 
discussed in literature in various contexts like the analysis of 
interferometric observations of the Cosmic Microwave Background Radiation 
\citep[e.g.][]{ho95}, the large-scale H~{\sc i} distribution at high redshifts 
\citep{bh01} and the interferometric H~{\sc i} observations to detect the epoch of reionization \citep{mo04,bh05}. The technical issues like the effect of the 
window function corresponding to the size of the source on the power spectrum 
estimator and the method of avoiding the noise bias by correlating the 
visibilities at two different baselines are described in detail in 
\citet{ba06}. The actual algorithm of estimating the power spectrum from the 
measured visibility function is outlined in \citet{dp08}. Here, a very similar 
algorithm, slightly modified to further reduce the noise bias, is used for the 
present work. To minimize the contribution of correlated noise power to the 
power spectrum estimator, visibilities are correlated at two different 
baselines with slightly different time-stamp for which the noise is expected 
to be uncorrelated. \citet{ba06} has shown that the real part of the measured 
visibility correlation directly estimates the power spectrum at baselines 
large compared to the inverse angular size of the source and at smaller 
baselines the true power spectrum is convolved with the window function. The 
error of the power spectrum is estimated accounting for both the noise in the 
measured visibility function and the finite number of independent estimates of 
the true power spectrum (cosmic variance).

\section{Data and Results}
\label{sec:dar}

\subsection{Summary of the data}

\begin{table}
 \caption{Details of the VLA and the GMRT data}
\begin{center}
 \begin{tabular}{lcccc}
 \hline
Source & Wavelength & Array & Project & Date of observation \\
\hline
\multicolumn{2}{c}{The VLA archival data:} & & &\\
\cline{1-2}
Cas A & 6 cm & A & AR0435 & 09 Dec., 2000\\
Cas A & 6 cm & A & AR0435 & 10 Dec., 2000\\
Cas A & 6 cm & B & AR0435 & 25 Mar., 2001\\
Cas A & 6 cm & B & AR0435 & 29 Apr., 2001\\
Cas A & 6 cm & C & AR0435 & 25 Apr., 2000\\
Cas A & 6 cm & D & AR0435 & 07 Sep., 2000\\
Crab  & 6 cm & A & AH0337 & 19 Oct., 1988\\
Crab  & 6 cm & A & AH0337 & 08 Nov., 1988\\
Crab  & 6 cm & B & AB0876 & 09 Aug., 1998\\
Crab  & 6 cm & C & AB0876 & 27 Jan., 1999\\
Crab  & 6 cm & D & AH0625 & 19 Nov., 1997\\
\cline{1-2}
\multicolumn{2}{c}{The GMRT data:} & & &\\
\cline{1-2}
Cas A & 20 cm & -- & 11NRb02 & 03 Dec., 2006\\
\hline
\end{tabular}\\
\end{center}
\label{table:obs}
\end{table}

The Giant Metrewave Radio Telescope \citep[GMRT;][]{gs91} L-band ($20$ cm) 
receiver was used to observe the supernova remnant Cas A. The unique hybrid 
array configuration of the GMRT allows one to probe structures on both large 
and small angular scale in a single observation. Scans on standard calibrators 
were used for flux calibration, phase calibration and also to determine the 
bandpass shape. The Very Large Array (VLA) archival C-band ($6$ cm) data are 
also used for both the supernova remnant Cas A and Crab Nebula. A summary of 
the GMRT and the VLA data used for this work with the observation band, the 
telescope array configuration, the original programme code and the dates of 
observation is given in Table (\ref{table:obs}). Data analysis was carried out 
using standard AIPS. After flagging out bad data, the flux density scale and 
instrumental phase were calibrated. The calibrated visibility data of the 
target sources are then used to estimate the angular power spectra and the 
errors.

\subsection{Results for Crab Nebula and Cas A}

\begin{figure}
\begin{center}
\includegraphics[scale=0.65, angle=0.0]{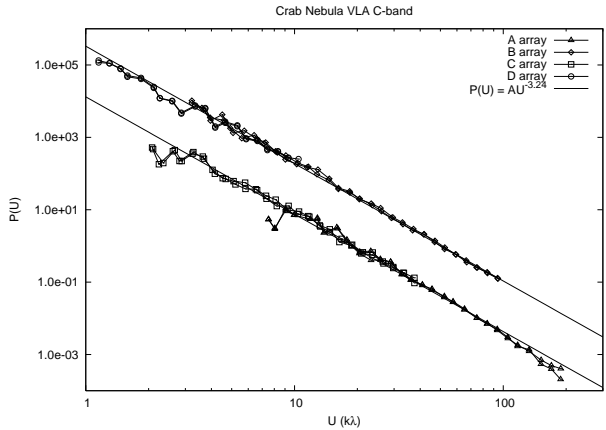}
\caption{\label{fig:1} \small{Intensity fluctuation power spectra for the Crab 
Nebula. The data points are from the VLA 6 cm (C-band) observation with 
different array configurations and at two different IFs. The line is the best 
fit power law with the power law index of $-3.24$. Data from different array 
configuration and the best fit power law are shown here with an offset in 
amplitude for clarity.}}
\end{center}
\end{figure}

\begin{figure}
\begin{center}
\includegraphics[scale=0.65, angle=0.0]{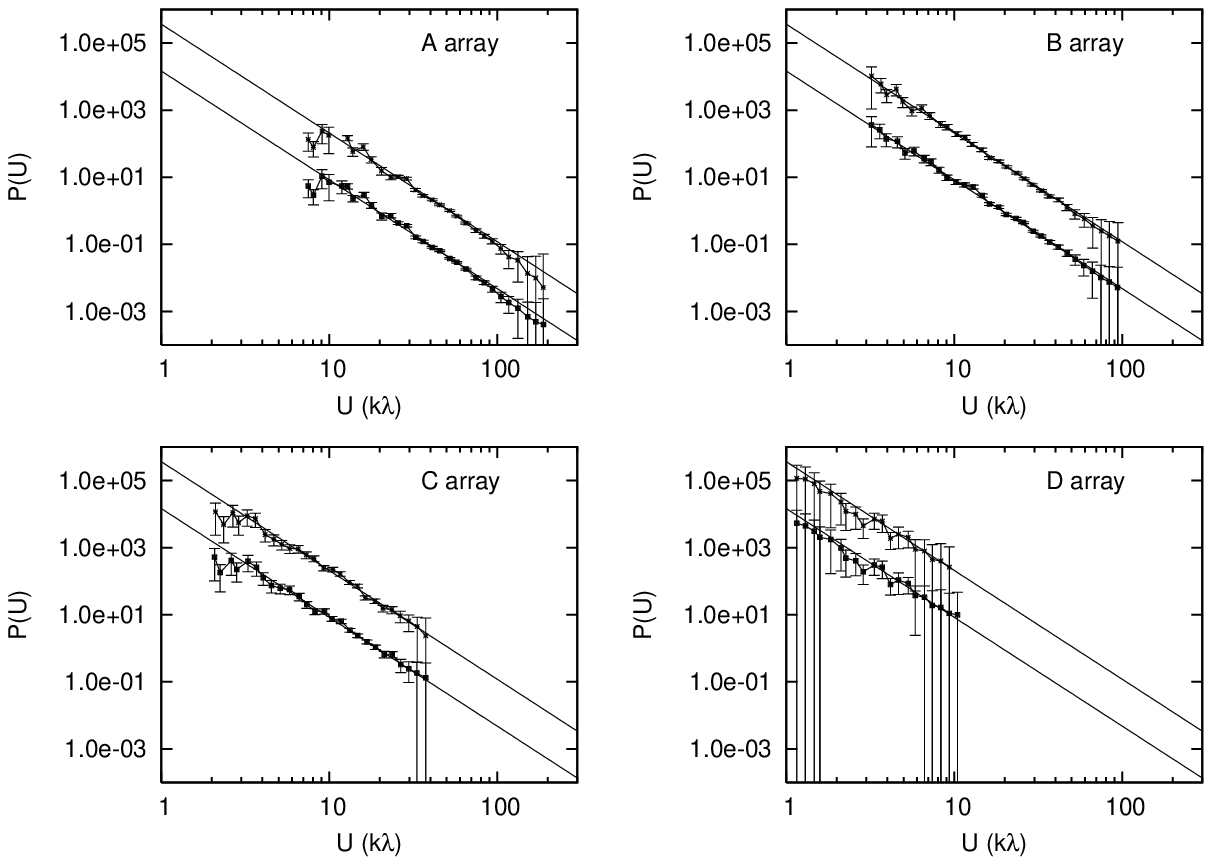}
\caption{\label{fig:2} \small{Intensity fluctuation power spectra with $\pm 1 
\sigma$ errorbars for the Crab Nebula. The VLA 6 cm (C-band) power spectra 
derived from the observation with different array configurations are shown in 
different panels. Results from two different IFs are plotted in the same panel 
with an offset in amplitude. The line is same as in Figure (\ref{fig:1}).}}
\end{center}
\end{figure}

\begin{figure}
\begin{center}
\includegraphics[scale=0.65, angle=0.0]{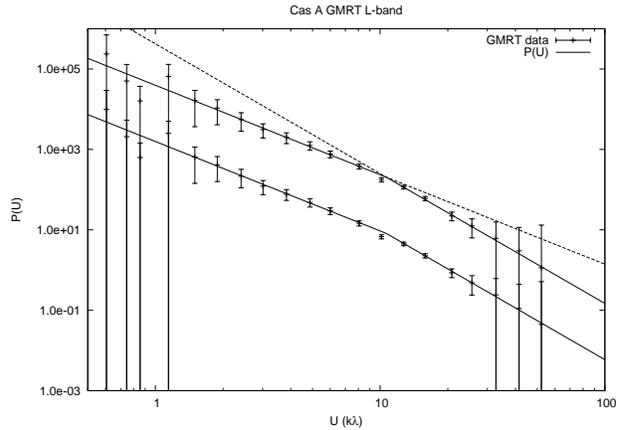}
\caption{\label{fig:3} \small{Intensity fluctuation power spectra for the 
supernova remnant Cas A. The points with $\pm 1 \sigma$ errorbars are from 
the GMRT 20 cm (L-band) data with two different observation frequencies. The 
line is showing the best fit power law with power law index of $-2.22$ and 
$-3.23$ before and after the break (at $10.6$ k$\lambda$) respectively. Power 
spectra derived from different frequency ranges and the best fit function are 
plotted with an offset in amplitude.}}
\end{center}
\end{figure}

\begin{figure}
\begin{center}
\includegraphics[scale=0.65, angle=0.0]{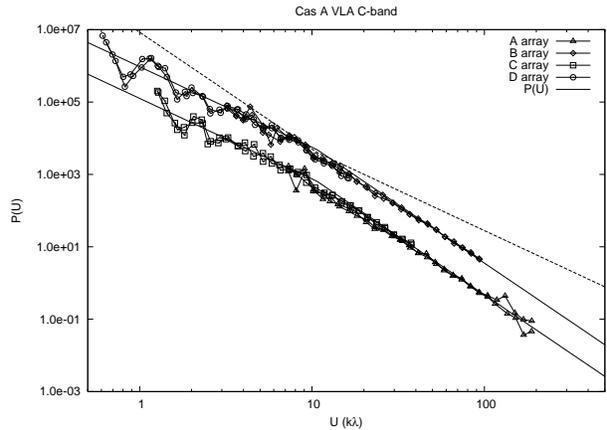}
\caption{\label{fig:4} \small{Intensity fluctuation power spectra for Cas A. 
The data points are from the VLA 6 cm (C-band) observation with different array 
configurations and at two different IFs. The line is the best fit function as 
in Figure (\ref{fig:3}). Data from different array configuration and the best 
fit function are shown here with an offset in amplitude.}}
\end{center}
\end{figure}

\begin{figure}
\begin{center}
\includegraphics[scale=0.65, angle=0.0]{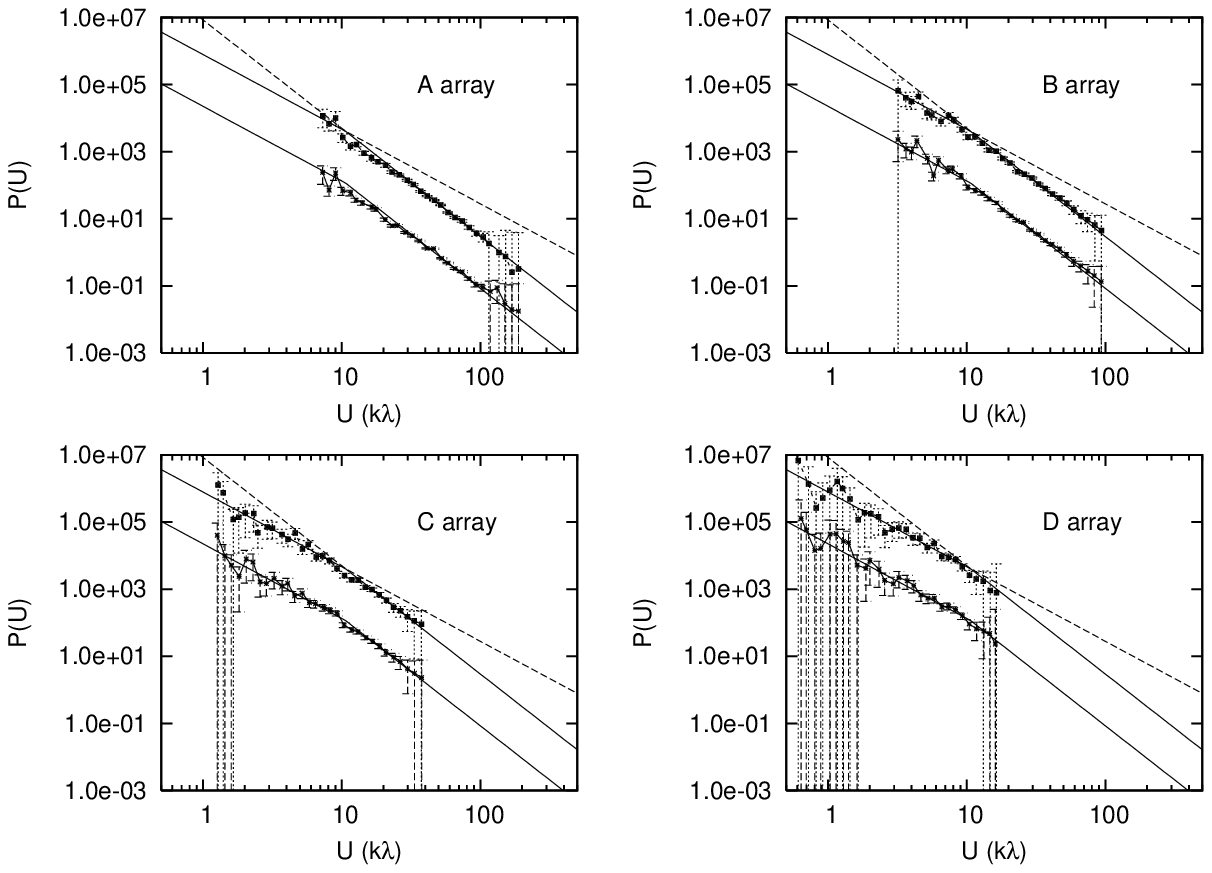}
\caption{\label{fig:5} \small{Intensity fluctuation power spectra with $\pm 1 
\sigma$ errorbars for Cas A. The VLA 6 cm (C-band) power spectra derived from 
the observation with different array configurations are shown in different 
panels. Results from two different IFs are plotted in the same panel with an 
offset in amplitude. The line is same as in Figure (\ref{fig:3}).}}
\end{center}
\end{figure}

For the Crab Nebula, the angular power spectrum as a function of inverse 
angular scale is found to be power law with a power law index of $-3.24 \pm 
0.03$. Since the effect of the convolution with the window function is more 
significant in the shorter baselines, the power law index is extracted by 
fitting the power spectra in the longer baseline range of $6-60$ k$\lambda$. 
Figure (\ref{fig:1}) shows the spectra derived from the VLA 6 cm observation 
with different array configuration and at two different intermediate 
frequencies (IFs) . The best fit power law, with an offset introduced in 
amplitude for clarity, is also shown in the same figure. The four panels in 
Figure (\ref{fig:2}) show the power spectra with $\pm 1 \sigma$ errorbars 
derived using the data from different VLA array configuration and IFs. The 
noise in the measured visibility function dominates at long baselines and the 
cosmic variance is the significant source of error at small baselines. It is 
clear from these figures that for a wide range of scales (about $1-100$ 
k$\lambda$ which corresponds to an angular scale of $2.5-250$ arcsec), the 
intensity fluctuation angular power spectra is a power law function of inverse 
angular scale. The power spectra derived from data from different array 
configurations and different IFs are in very good agreement.

The power spectrum for Cas A is also found to be a power law with a very 
similar power law index at small angular scales. But, as shown in Figure 
(\ref{fig:3}), there is a break in the spectrum at about $10.6$ k$\lambda$ 
($\sim 25$ arcsec) and the power law index changes significantly at smaller 
$U$. The best fit power law function for the power spectrum derived from the 
GMRT 20 cm data has a power law index of $-2.22\pm0.03$ in the shorter 
baseline range ($1.6-10$ k$\lambda$). After the break the index changes to 
$-3.23\pm0.09$ (estimated from the range $11-30$ k$\lambda$) and the power 
spectrum remains steeper all the way up to the smallest angular scale ($\sim 
5$ arcsec) probed in this observation. These results for Cas A power spectrum 
are consistent with the power spectra derived from the VLA 6 cm archival 
data of Cas A. It is found that the 6 cm power spectrum is also a broken 
power law with the same power law index and the break at the same angular 
scale as in the 20 cm power spectrum. The VLA 6 cm power spectra obtained 
from observation with different array configurations and two different IFs are 
plotted in Figure (\ref{fig:4}). This shows that the steeper power law at the 
long baseline range is in fact extended upto $100$ k$\lambda$ ($\sim 2.5$ 
arcsec). The four panels in Figure (\ref{fig:5}) show the power spectra with 
$\pm 1 \sigma$ errorbars derived using the data from four different VLA array 
configuration and two IFs. Clearly, for Cas A also, the power spectra derived 
from 20 cm and 6 cm data with different array configurations and different 
IFs are in good agreement. 

\subsection{Interpretation of the results}

The power law index of the steeper part of the Cas A angular power spectrum is 
completely consistent, within the measurement errorbars, with the power law 
index of the Crab Nebula power spectra. The break in the Cas A power spectrum 
and the change of the power law index at small baseline range (or large 
angular scale) is very interesting. We have verified analytically that the 
shell type geometry of Cas A will affect the power spectrum significantly only 
at very small $U$ by convolving it with a window function which is the Fourier 
transform of the two dimensional projection of this optically thin shell. The 
same is also true for the optically thin spherical geometry of the Crab Nebula. For 
the long baseline range around $10$ k$\lambda$, the effect will be negligible 
and can not explain the sharp break and the significant change of power law 
index by $\sim 1$. It appears that a plausible explanation is a transition 
from  three dimensional at small scales ($U > 10$ k$\lambda$) to two 
dimensional turbulence at large scales ($U < 10$ k$\lambda$). The shell 
thickness sets the angular scale of the transition. On length scales smaller 
than the shell thickness, the shell can have modes of perturbation in all 
three independent directions. But on length scales larger than the shell 
thickness, there will be no modes perpendicular to the shell thickness. This 
makes the turbulence to change from a three dimensional to effectively a two 
dimensional in nature and hence the power law index changes by $1$. This 
change in slope may  possibly be related to the fact that the slope of the 
velocity power spectrum changes from $-11/3$ to $-8/3$ in going from 3D to 2D 
for incompressible, Kolmogorov turbulence \citep{ko41}. The density power 
spectrum is predicted to follows the velocity power spectrum in the 
\citet{gs95} model of MHD turbulence. The observation, that the angular scale 
of this break matches approximately with the shell thickness, is indicative of 
the consistency of this picture. A similar difference of $\approx 1$ in the 
power law index has also been observed and interpreted as a transition from 
three dimensional turbulence to two dimensional turbulence in the power 
spectrum of H~{\sc i} 21 cm emission intensity fluctuations of the Large 
Magellanic Cloud \citep{el01} and the galaxy NGC~628 \citep{dp08}. 

The scale-free nature of the power spectra over a wide range of scales and a 
very similar value of the power law index for power spectra of two very 
different type of supernova remnants suggests the universality of the physical 
process responsible for the observed intensity fluctuation. We propose that 
the fluctuation is most probably due to the turbulence in the synchrotron 
emitting plasma that gives rise to the power law power spectrum. The 
interaction of the propagating shock with the turbulent interstellar medium is 
known to enhance the turbulence in the postshock region and causes the spatial 
variation of emission in supernova remnants \citep{ba01}.  We investigate here 
whether the observed power spectrum $P(k) \propto k^{-3.2}$, or equivalently 
the energy spectrum $E(k) = k^2 P(k) \propto k^{-1.2}$, is consistent with our 
present understanding of astrophysical turbulence. The observed intensity 
fluctuation power spectra is related to the density and magnetic field power 
spectra which, in turn, are found, from numerical simulations, to closely 
follow the velocity fluctuation power spectra. For incompressible and 
nonmagnetized turbulence Kolmogorov theory suggests an isotropic power law 
velocity fluctuation energy spectrum $E_v(k)\propto k^{-5/3}$ where $k$ is the 
magnitude of the wave vector \citep{ko41}. \citet{ir64} and \citet{kr65} gave 
a model of magnetic incompressible turbulence (IK theory) that predicts, even 
in the presence of magnetic field, isotropic power law energy spectra $E(k) 
\propto k^{-3/2}$ for both velocity and magnetic field. Without any assumption 
of isotropic energy distribution, \citet{gs95} proposed a model of 
incompressible magnetohydrodynamic turbulence that predicts a Kolmogorov-like 
energy spectra $E_v(k_\perp) \propto k_\perp^{-5/3}$ where $k_\perp$ is the 
component of the wave vector perpendicular to the local magnetic field 
direction. It also predicts an anisotropy condition $k_\parallel \propto 
k_\perp^{2/3}$ where $k_\parallel$ is the component of the wave vector 
parallel to the local magnetic field direction. But, even if there is 
anisotropy in the system of reference defined by the local magnetic field, it 
is worth keeping in mind that there will only be moderate anisotropy in the 
observer's reference. 

For compressible magnetohydrodynamics turbulence, there is, unfortunately, no 
widely-accepted theory and much of the present understanding has come from 
numerical results. Recent numerical simulation indicates that, for 
compressible magnetohydrodynamic turbulence, both the velocity and magnetic 
field energy spectra and anisotropy in Alfv$\acute{e}$n modes and slow modes 
are as predicted by \citep{gs95}. But the energy spectra for fast modes are 
isotropic and the scaling is as predicted in IK theory \citep{cl02}. It is 
also found that, at least in case of incompressible magnetic turbulence, 
viscous damping on scales larger than the magnetic diffusion scale can make 
the magnetic energy spectrum significantly less steep. \citet{ch02} reports 
magnetic energy spectrum $E_b(k) \propto k^{-1}$ implying rich structure of 
magnetic field on small scales. 

The synchrotron emissivity $i_s \propto n_e|B_\perp|^{(p+1)/2}$ where $n_e$ is 
the electron number density, $B_\perp$ is the magnetic field component 
perpendicular to the line of sight and the typical value of the power law 
index $p$ of electron energy distribution in supernova remnants is about $2$ 
\citep{gr91}. One dimensional numerical analysis 
suggests that if the magnetic field power spectrum is a power law, then 
$|B|^{(p+1)/2}$ will also have a power law spectrum with the same index for 
the values of $p$ in the range of our interest. But, because of this 
nonlinearity, in general it is not straightforward to derive the magnetic 
field fluctuation power spectrum from the intensity fluctuation power 
spectrum. It is also not necessarily true that the density distribution and 
magnetic field are strongly coupled. Numerical simulation and analytical study 
in fact suggest that in compressible magnetohydrodynamic turbulence, the 
magnetic field strength and density are only weakly correlated \citep{pa03}. 
\citet{be05} also reports a flat and isotropic density spectrum from numerical 
simulation of supersonic magnetohydrodynamic turbulence. But, if the electron 
density distribution smoothly follows the magnetic field inhomogeneities in 
the supernova remnants, the synchrotron intensity fluctuation power spectrum 
is directly related to the magnetic field power spectrum. In this condition, 
following the analysis of \citet{ddg00}, one can conclude that if the 
intensity fluctuation has a power law power spectrum then the magnetic field 
fluctuation will also have a power law spectrum with the same power law index, 
provided that the magnetic field perturbation amplitude is small. The effect 
of the nonlinear law of synchrotron emission on the power spectrum is not 
clear in situations when the perturbation amplitude is high or correlation 
between the magnetic field strength and density is weak. 

The observed intensity fluctuation power spectrum is somewhat shallower than 
the expected spectrum of the magnetic field. From the above discussion, one 
can identify three plausible reasons for this discrepancy which can make the 
spectrum less steep. They are (i) viscous damping on scales larger than the 
magnetic diffusion scale, (ii) weak or no correlation between the magnetic 
field and the density distribution and (iii) large amplitude of magnetic field 
perturbation which may affect via the nonlinearity of synchrotron emissivity. 

Interestingly, numerical, observational and theoretical studies of synchrotron 
emission fluctuations are carried out in a completely different context to 
understand the effect of the Galactic foreground emission on the angular power 
spectrum of the cosmic microwave background \citep[see][and references therein 
for details]{cl02a} and in some of the cases the intensity fluctuations are 
attributed to turbulence \citep{ch98,te00,cl02a}. Though in a very different 
range of angular scale, the energy spectrum of the Galactic synchrotron 
foreground is found to be a range of power laws with power law index $\sim-1$ 
for higher latitudes \citep[][and references therein]{cl02a}. This is less 
steep than the expected $k^{-5/3}$ energy spectrum of the magnetic field. 
Clearly, a similar discrepancy is evident in this case also. But, in spite of 
this discrepancy, one can say that the near-Kolmogorov power law power 
spectrum is broadly consistent with our present understanding of the 
magnetohydrodynamic turbulence. 

\section{Conclusions}
\label{sec:con}

We have analysed the data from GMRT 20 cm and VLA 6 cm observations of two 
supernova remnants Cas A and Crab Nebula and estimated the angular power 
spectra of the synchrotron radiation intensity fluctuation over a wide range 
of angular scale. We report, for the first time, a power law power spectrum of 
the synchrotron radiation intensity fluctuation in supernova remnants. The 
power law index is found to be $-3.24\pm0.03$ for both these sources with very 
different large scale morphology. For Cas A, there is a break in the power 
spectrum and the power law index changes from $-3.2$ to $-2.2$ at angular 
scale larger than the size of the shell thickness. This change is a result of 
the anisotropy of the perturbation at length scales larger than the shell 
thickness. This power law power spectrum is consistent with our present 
understanding of the magnetohydrodynamic turbulence derived mostly from 
existing numerical simulation results. 

\section*{Acknowledgments}

This research has made use of the NASA's Astrophysics Data System, the 
National Radio Astronomy Observatory (NRAO) VLA archival data and data from 
the GMRT observation. The NRAO is a facility of the National Science Foundation 
operated under cooperative agreement by Associated Universities, Inc. We thank 
the staff of the GMRT who have made these observations possible. The GMRT is 
run by the National Centre for Radio Astrophysics of the Tata Institute of 
Fundamental Research. NR would like to acknowledge the hospitality of all the 
staff members of the Centre for Theoretical Studies (Indian Institute of 
Technology, Kharagpur) during his stay for this collaboration.We are grateful 
to Dipankar Bhattacharya, Dave Green, Ranjeev Misra, Rajaram Nityananda, 
Dmitri Pogosyan, A. Pramesh Rao and Kandaswamy Subramanian for much 
encouragement and many helpful comments. We are also grateful to the anonymous 
referee for useful comments.

\bsp

\label{lastpage}

\end{document}